\documentclass[aps,prb,preprintnumbers,amsmath,amssymb,superscriptaddress,twocolumn,10pt]{revtex4-2}
\usepackage{txfonts}
\usepackage{graphicx}
\usepackage{dcolumn}
\usepackage{bm}
\usepackage{array}
\usepackage[dvipsnames]{xcolor}
\usepackage[%
    colorlinks=true,
    pdfborder={0 0 0},
    linkcolor=blue
]{hyperref}  
\usepackage{chngpage}
\usepackage{appendix}
\usepackage{subfig}
\usepackage{booktabs}
\begin{document}
\graphicspath{ {figuresPy/} }

\def\V {$^{51}$V }
\def\Sb {$^{121}$Sb }
\def\Rb {$^{87}$Rb }

\title{Microscopic nature of the charge-density wave in kagome superconductor RbV$_3$Sb$_5$}

\author{\hspace{1mm}Jonathan Frassineti}
 
 \affiliation{
Dipartimento di Fisica e Astronomia, Universit\'a di Bologna, I-40127 Bologna, Italy }
 
\author{\hspace{1mm}Pietro Bonf\`a}
\affiliation{
Dipartimento di Scienze Matematiche, Fisiche e Informatiche, Universit\'a di Parma, I-43124 Parma, Italy
}

\author{\hspace{1mm}Giuseppe Allodi}
\affiliation{
Dipartimento di Scienze Matematiche, Fisiche e Informatiche, Universit\'a di Parma, I-43124 Parma, Italy
}

\author{\hspace{1mm}Erick Garcia}
\affiliation{
Department of Physics, Brown University, Providence, Rhode Island 02912, USA}

\author{\hspace{1mm}Rong Cong}
\affiliation{
Department of Physics, Brown University, Providence, Rhode Island 02912, USA}

\author{\hspace{1mm}Brenden R. Ortiz} 
\affiliation{
Materials Department and California Nanosystems Institute, University of California Santa Barbara,
Santa Barbara, California 93106, USA}

\author{\hspace{1mm}Stephen D. Wilson}
\affiliation{
Materials Department and California Nanosystems Institute, University of California Santa Barbara,
Santa Barbara, California 93106, USA}

\author{\hspace{1mm}Roberto De Renzi}
\affiliation{
Dipartimento di Scienze Matematiche, Fisiche e Informatiche, Universit\'a di Parma, I-43124 Parma, Italy
}

\author{\hspace{1mm}Vesna F. Mitrovi{\'c}}
\affiliation{
Department of Physics, Brown University, Providence, Rhode Island 02912, USA}

\author{\hspace{1mm}Samuele Sanna}
 \affiliation{
Dipartimento di Fisica e Astronomia, Universit\'a di Bologna, I-40127 Bologna, Italy }

\date{\today}
\begin{abstract}
The recently discovered vanadium-based kagome metals AV$_3$Sb$_5$ (A = K, Rb, Cs) offer the possibility to study the interplay between competing electronic orderings, such as charge density order and superconductivity. We focus on the former and provide a comprehensive set of $^{51}$V, $^{87}$Rb, and $^{121}$Sb  magnetic resonance measurements on an RbV$_3$Sb$_5$ single crystal. Elucidating the symmetries and properties of the CDW phase is essential to understanding the unconventional electronic orderings occurring in this material. We establish the structure of the $2\times 2 \times 2$ superlattice that describes the system below the charge density wave transition by combining both experimental and computational methods, with a methodology that can be readily applied to the remaining compounds of the same family. Our results give compelling evidence that the CDW structure occurring below 103 K for RbV$_3$Sb$_5$ is the so-called Inverse Start of David pattern $\pi$-shifted along the c axis (also known as staggered tri-hexagonal). 
\end{abstract}

\maketitle

Over the last years, kagome materials have attracted widespread interest in the field of condensed matter due to the possibility of studying the ground and excited states that emerge from the interplay between a frustrated geometry of the crystalline structure and a nontrivial band topology \cite{Depenbrock2012, Ghimire2020,Huse1992,Kitaev2006,Koshibae2003,Li2018,Morita2018,Singh2008,Taillefumier2014,Yamada2016, Kiesel2013,Wang2013,Yu2012}.
In this context, the vanadium-based kagome materials AV$_3$Sb$_5$ (A = Rb, Cs, K) \cite{wang2020proximityinduced,PhysRevMaterials.3.094407,PhysRevLett.125.247002,Yang2020} are of particular interest owing to the recent discovery of a superconducting ground state below $T_c$ $\sim$ 0.9 - 2.5 K.
In addition to the low temperature superconducting transition, these systems undergo a Charge Density Wave (CDW) transition at a temperature $T_{CDW}$ $\sim$ 80 - 104 K \cite{song2021orbital,PhysRevLett.125.247002,PhysRevMaterials.3.094407,shumiya2021tunable}. ARPES combined with DFT suggests that the specific order of the CDW is cation-dependent \cite{kang2022}. Scanning tunneling microscopy (STM) and Muon spin spectroscopy ($\mu$SR) observed a chiral charge order that breaks time-reversal
symmetry \cite{Jiang2021,shumiya2021tunable,oey2022fermi,guguchia2022}, which leads to the anomalous Hall effect even in the absence of magnetism of electronic origin \cite{Graf2021,Yang2020}.
The CDW order plays a central role in the definition of the superconducting gap structure \cite{PhysRevB.105.L100502} and has therefore been a matter of great attention. 

\begin{figure*}[ht]
\centering
\includegraphics[width =  \linewidth]{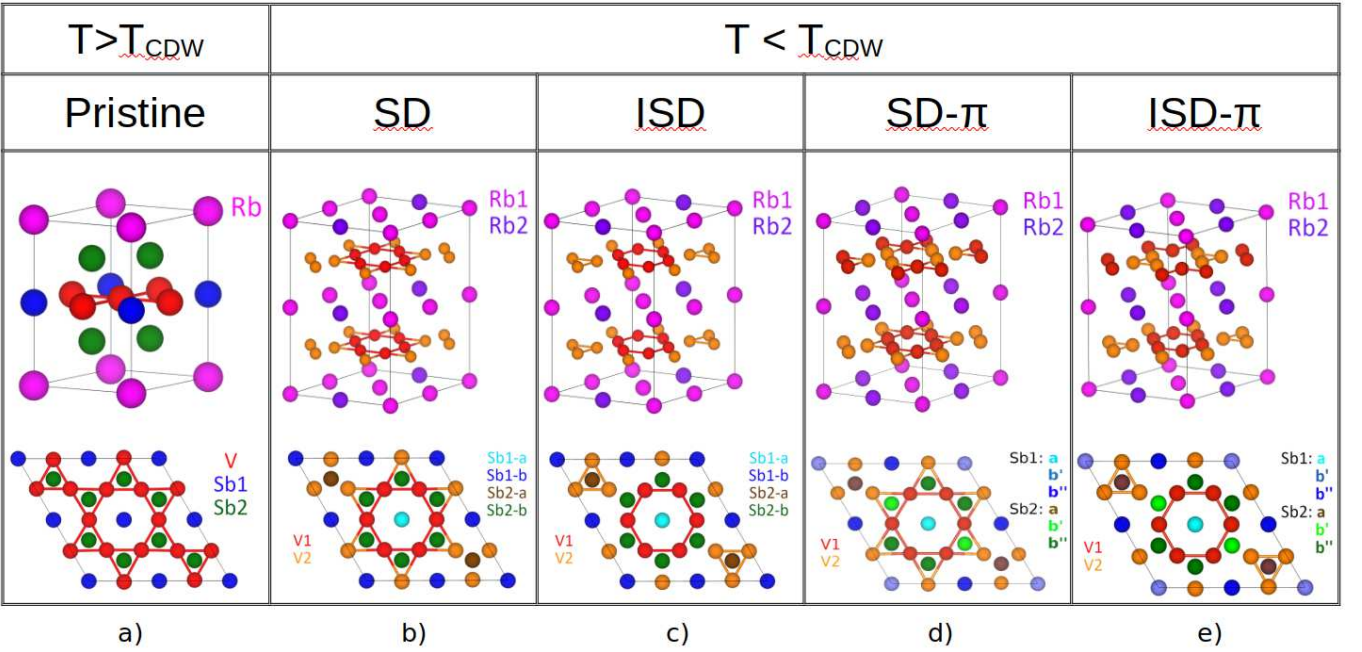}
\caption{(Color online) The crystalline structure of RbV$_{3}$Sb$_{5}$, above and below $T_{CDW}$. a) The pristine phase adopted above $T_{CDW}$, with the V atoms (in red) forming a 2D kagome lattice. The Sb atoms are divided into two sets: Sb1 (in blue) and Sb2 (in dark green). The Rb atoms are labeled in magenta. b), c), d), e) The four possible structures proposed below $T_{CDW}$: Star-of-David (SD), Inverse Star-of-David (ISD) or tri-hexagonal, (staggered) Star-of-David and Inverse Star-of-David with $\pi$-shift (SD-$\pi$ and ISD-$\pi$, respectively). Here, the Sb atoms in the top figures are not displayed for clarity purposes.}
\label{fig:crystal}
\end{figure*}

In this work, we focus on RbV$_{3}$Sb$_{5}$ and unambiguously identify the ground state structure below the CDW transition.
At room temperature, RbV$_{3}$Sb$_{5}$ has a layered structure (henceforth Pristine) structure with hexagonal symmetry (space group P6/mmm, No. 191), as shown in Fig.~\ref{fig:crystal}a.
It consists of Rb layers and V-Sb
slabs alternated along the $c$ axis, and it is isostructural to KV$_{3}$Sb$_{5}$ and CsV$_{3}$Sb$_{5}$ \cite{PhysRevMaterials.3.094407}. 
The most important feature of this material is the two-dimensional (2D) kagome layer formed by the V atoms (Fig.~\ref{fig:crystal}a), whose $3d$ orbitals give rise to a number of peculiar features in the electronic band structure extensively described in literature \cite{Hu2022,song2021orbital}.
One of the key features is the occupation of van Hove singularities close to the Fermi energy, and the Sb p-orbitals have also been suggested to play a crucial role in completing the description of the three Van Hove singularity structures \cite{PhysRevB.105.235145}.

For $T>T_{CDW}$, the high-temperature phase contains two inequivalent Sb sites: Sb1 sites are co-planar with V, while Sb2 sites are located off-plane below and above the V layer. Sb1 sits in the center of V hexagons while Sb2 prospectively falls in the center of V triangles, forming graphene-like hexagon layers. 

In the CDW state, below $T_{CDW}$ = 103 K, the lattice undergoes a structural transition.
First-principles calculations initially proposed two possible distortions for the arrangement of the kagome planes. One is the Star-of-David (SD) distortion (Fig.~\ref{fig:crystal}b) of V atoms, which resembles the well-known motif of the CDW state found in transition-metal dichalcogenides \cite{doi:10.1080/00018737500101391,Rossnagel_2011}. The other is an inverse deformation of the SD pattern (ISD, Fig.~\ref{fig:crystal}c). This results in a periodic arrangement of V atoms in triangles and hexagons which is also called Tri-Hexagonal (TrH) structure. 

Regarding the stacking along the c axis, different arrangements have been proposed: i) repetition of SD or ISD (Fig.~\ref{fig:crystal}b and c), ii) alternation of ISD and SD, iii) a $\pi$-shift translation between adjacent planes (SD-$\pi$ and ISD-$\pi$, Fig.~\ref{fig:crystal}d and e). 
The overall distortion must therefore be described in, at least, a 2$\times$2$\times$2 supercell.
Higher order modulations have been observed in the Cs variant only, but they manifest in a competing state nearly degenerate with a 2x2x2 phase.

Table \ref{tab:structures} summarizes the currently proposed CDW structures, as they are identified by different techniques for each compound of the family, showing that the attribution is still highly controversial and possibly material dependent.
Actually, the identification of the precise CDW configuration and symmetries is crucial to give the proper basis for theoretical modeling of the exotic orders predicted for this family compound \cite{arxiv.2207.12820}.

\begin{table}[htbp]
\caption{
Charge-density wave structures ($T < T_{CDW}$) for AV$_{3}$Sb5 (A = Cs, K, Rb) from the literature.
}\label{tab:structures}
\begin{tabular}{|c|c|c|c|}
\hline
\hline
& CsV$_{3}$Sb$_{5}$  & KV$_{3}$Sb$_{5}$  & RbV$_{3}$Sb$_{5}$ \\
\hline

ISD-$\pi$   &  N \cite{song2021orbital, nie2022}, X \cite{PhysRevX.11.031050, PhysRevB.105.195136}     &  A+D \cite{ kang2022, tan2021charge} &    X \cite{PhysRevX.11.031050}, D \cite{kang2022, tan2021charge}  \\
&  S \cite{PhysRevX.11.031026}, D \cite{tan2021charge} & A \cite{jiang2022observation} & \\
\hline
SD-$\pi$  & N \cite{song2021orbital}, X \cite{PhysRevX.11.031050, PhysRevB.105.195136} &  /  &  X \cite{PhysRevX.11.031050} \\
& S \cite{PhysRevX.11.031026} & & \\
\hline
SD+ISD &  X \cite{OrtizPhysRevX2021}, A+D \cite{arxiv.2201.06477, kang2022}  & A+D \cite{arxiv.2201.06477} & A+D \cite{arxiv.2201.06477}   \\
\hline
SD & N \cite{luo2022} & / & / \\
\hline
ISD & N \cite{Mu_2022} & A \cite{luoHailan2022, Kato2022} & A \cite{PhysRevLett.127.236401} \\
\hline
\hline
\multicolumn{4}{|c|}{Legend: N = NMR, X = XRD, S = STM, A = ARPES, D = DFT} \\
\hline
\hline
\end{tabular}
\end{table}

In this work, we resolve this controversy for RbV$_{3}$Sb$_{5}$ by combining $^{51}$V/$^{87}$Rb NMR and $^{121}$Sb NQR measurements with Density Functional Theory (DFT) calculations unambiguously identifying the crystalline structure of the CDW phase.


To this aim, $^{51}$V/$^{87}$Rb nuclear magnetic resonance (NMR) spectra have been collected on a single crystal of RbV$_{3}$Sb$_{5}$ as a function of temperature, with an external magnetic field $B_0$= 7.95 T and 6.99 T. In addition, $^{121}$Sb zero-field nuclear quadrupolar resonance (NQR) spectra have also been acquired.
The experimental procedure is detailed in SM \footnote{See Supplemental Material at [URL will be inserted by publisher] for NMR/NQR methods on RbV$_{3}$Sb$_{5}$ single crystal and computational details. The Supplemental Material includes Refs.~\cite{PhysRevLett.125.247002, Yin_2021,Clark1995,elk,PhysRevB.40.3616,PhysRevLett.77.3865,PhysRevB.13.5188,PhysRevB.16.1748,tan2021charge}. All files related to a published paper are stored as a single deposit and assigned a Supplemental Material URL. This URL appears in the article’s reference list.}.

In the NMR/NQR spectra, the resonance frequency is sensitive to the magnetic and charge environment, hence non-equivalent nuclear sites arising from the symmetry breaking effects are typically detected by frequency peak splittings across the phase transition. 
In addition, the area $A_i$ of the $i-th$ NMR/NQR peak is proportional to the statistical occupation of a specific $i-th$ nucleus, i.e. to the multiplicity of each non-equivalent site in the unit cell.
We base the determination of the CDW configuration upon:  i) the $A_i/A_j$ ratio of the NMR/NQR peaks for each couple of non-equivalent nuclei ($i \neq j$), equal to the population ratio of the non-equivalent nuclear site per species and ii) on the quadrupolar frequency $\nu_{Q}$ and asymmetry parameter of the Electric Field Gradient (EFG) tensor $\eta$ of each peak that we compare to DFT based calculations of the EFG in the various phases \footnote{The detailed description of the computational approach is provided in the SM ~\cite{Note1}.}.

In the lowest symmetry configurations considered here, i.e. SD-$\pi$ and ISD-$\pi$, the V and Rb atoms split into two sublattices, while three non-equivalent sites are obtained for Sb1 and Sb2 nuclei.
These are shown in Fig.~\ref{fig:crystal}d and e. 
The multiplicity of each non-equivalent site is summarised in Table ~\ref{tab:occupations} for both the pristine ($T > T_{CDW}$) and various proposed CDW ($T < T_{CDW}$) phases.

\begin{table}[htbp]
\caption{
Atomic occupations for \V, \Sb and \Rb atoms in RbV$_{3}$Sb$_{5}$ above and below CDW transition.
}\label{tab:occupations}
\begin{tabular}{|c|c|c|}
\hline
\hline
Phase & Atoms & Multiplicity  \\
\hline
\hline
 &   Rb & 1    \\
Pristine & V & 3\\
& Sb1 & 1 \\
& Sb2 & 4\\
\hline
&   Rb1/Rb2 & 6/2    \\
2$\times$2$\times$2 CDW  &   V1/V2 & 12/12   \\
(ISD or SD, \textcolor{red}{no} $\pi$-shift)&   Sb1-a/Sb1-b & 2/6    \\
&   Sb2-a/Sb2-b & 8/24    \\
\hline
 &   Rb1/Rb2 & 4/4    \\
2$\times$2$\times$2 CDW &   V1/V2 & 12/12   \\
 (ISD or SD, \textcolor{red}{with} $\pi$-shift)&   Sb1-a/Sb1-b'/Sb1-b'' & 2/2/4    \\
&   Sb2-a/Sb2-b'/Sb2-b'' & 8/8/16    \\
\hline
\hline
\end{tabular}
\end{table}



$^{121}$Sb (with nuclear spin $I$ = 5/2) NQR measurements are the most sensitive to distinguish between the SD and ISD structures, as already shown for the case of CsV$_{3}$Sb$_{5}$ \cite{Mu_2022,nie2022}. 
Due to the different and lower symmetry of their position, Sb2 and Sb1 nuclei are distinguished into Sb2-a/Sb2-b and Sb1-a/Sb1-b when the SD or the ISD structures are considered, and further separated into Sb2-a/Sb2-b/Sb2-b'' and Sb1-a/Sb1-b'/Sb1-b'' non-equivalent sites when a $\pi$-shift is introduced, as shown in Fig.~\ref{fig:crystal}. 

\begin{figure}[htbp]
\centering
{\includegraphics[width = \linewidth]{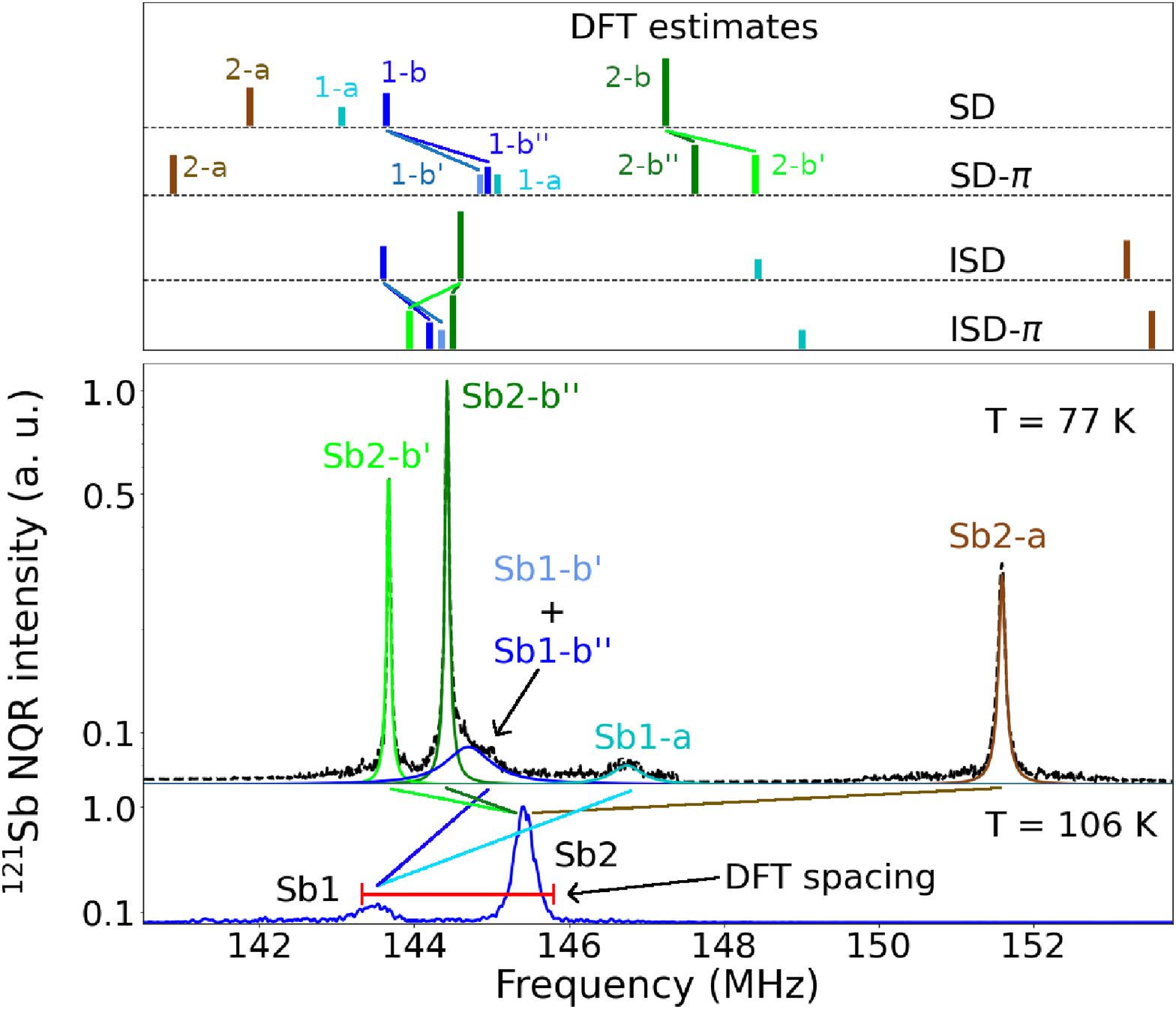}}
\caption{ Representative $^{121}$Sb NQR spectra above (106 K, bottom) and  below $T_{CDW}$ (77 K, center) for 5/2 $\rightarrow$ 3/2 NQR transition . In the top part, vertical ticks display the relative values calculated below $T_{CDW}$ by DFT for SD, ISD, and ISD-$\pi$ configurations with respect to the quadrupolar frequency of Sb sites above $T_{CDW}$ (see text); the height of the ticks is proportional to the site multiplicity. In the center, the $y$ scale is logarithmic. In the bottom part, the DFT spacing is calculated for the normal state.}
\label{fig:121Sb}
\end{figure}
Fig.~\ref{fig:121Sb} shows selected $^{121}$Sb NQR spectra for 5/2 $\rightarrow$ 3/2 NQR transition. 
Above $T_{CDW}$ = 103 K (bottom of Fig.~\ref{fig:121Sb}) two peaks are present and assigned according to the 1:4 occupation ratio Sb1:Sb2 (Fig.~\ref{fig:crystal}a and Table~\ref{tab:occupations}). 
In the CDW phase (center of Fig.~\ref{fig:121Sb}), we expect four peaks for a simple stacking of SD or ISD ordered kagome planes, while six peaks should appear when introducing a further $\pi$-shift along $c$. The data for $T < T_{CDW}$ show 5 peaks, with 4 of them clearly visible and an additional spectral weight appearing as a bump next to the most intense peak.
From this, we can exclude the two simpler SD or ISD structures with no pi-shift, since they would give only four peaks in the $^{121}$Sb NQR spectrum. 

According to the site multiplicity summarised in Table \ref{tab:occupations} the most intense peak, at almost 145 MHz, is assigned to the Sb2-b''.
To proceed further we estimate the quadrupolar frequencies for the various structures with ab initio simulations by collecting the electric field gradient (EFG) at each $^{121}$Sb nucleus in the SD, ISD, SD-$\pi$ and ISD-$\pi$ structures. As detailed in SM~\cite{Note1}, the absolute values of the calculated $\nu_{Q}$ at the Sb sites are affected by a systematic error, but these are canceled when the \textit{variations} of $\nu_{Q}$ between the pristine phase ($T>T_{CDW}$) and the low-temperature structures are considered.
The mismatch between the experimental results and the prediction for the SD structure is immediately evident (Sb2-a and Sb2-b nuclei have very different quadrupolar resonance frequencies in the SD and ISD structures), and therefore the addition of a $\pi$-shift for the SD structure does not improve the matching. Instead, a good agreement is found instead for the ISD structure and the small change introduced by the $\pi$-shift along the $c$ direction further improves the agreement between the prediction and the experiment. We can therefore safely rule out the presence of SD distortions in the kagome plane.



Table ~\ref{tab:areas} summarizes the $A_i/A_j$ ratio obtained from the area of the NQR peaks and the relative site occupation in the ISD-$\pi$ phase. 
\begin{table}[h!]
\caption{
Multiplicity ratio between $^{121}$Sb sites in the CDW phase, according to the crystalline structure (center) and experimental NQR area (right column).
}\label{tab:areas}
\begin{tabular}{|c|c|c|c|}
\hline
\hline
Phase & Sites & Crystal & NQR area (T = 77 K) \\
\hline
\hline
Pristine & Sb2:Sb1 &  4 & 4.20 $\pm$ 0.25              \\
\hline
& Sb2-b'':Sb1-a   &     8 & 7.77 $\pm$ 0.29          \\
CDW & Sb2-b':Sb1-a   &     4 & 4.02 $\pm$ 0.17   \\
 & Sb2-a:Sb1-a   &     4 & 3.87 $\pm$ 0.21   \\
& (Sb1-b''+Sb1-b'):Sb1-a   &     3 & 3.01 $\pm$ 0.18   \\
\hline
\hline
\end{tabular}
\end{table}
The values are in very good agreement, within the experimental uncertainty. As a result, we can attribute the bump at around 145 MHz to the combined signal from Sb1-b' and Sb1-b'' sites of the $\pi$-shifted ISD structure, as shown in center of Fig.~\ref{fig:121Sb}.

Incidentally, we note that the two peaks at the lowest and highest frequencies have been previously assigned to the Sb1-b and Sb2-a sites by earlier NQR studies \cite{Mu_2022,luo2022}. However, this conclusion is inconsistent with the expected ratio between the area of the two peaks.

To further justify the site assignment, we also collect the $3/2\rightarrow1/2$ NQR transition and we estimate the asymmetry parameter $\eta$ of the EFG tensor from  
the following ratio, valid if $\eta$ $\leq$ 0.1~\cite{AbragamBook,DasHahnBook,Note1}
\begin{equation}
    \frac{\nu_{Q}^{121} (5/2 \rightarrow 3/2)}{\nu_{Q}^{121} (3/2 \rightarrow 1/2)} \simeq 
    2(1 - \frac{35}{27}\eta^{2})\,.
\end{equation}

The experimental values of $\eta$ for Sb sites are compared with the DFT prediction for the ISD-$\pi$ structure in Table \ref{tab:eta_121_123}.
\begin{table}[h!]
\caption{
Comparison between experimental and predicted values for the asymmetry parameter $\eta$.
}\label{tab:eta_121_123}
\begin{tabular}{|c|c|c|c|}
\hline
\hline
Site & DFT (ISD-$\pi$) & Exp. (T = 77.3 K)\\
\hline
\hline
Sb1-a &      0 &  0.001 $\pm$ 0.005 \\
Sb1-b' &      0.056 &  0.034 $\pm$ 0.005  \\
Sb1-b'' &      0.058 &  0.034 $\pm$ 0.005  \\
Sb2-a   &      0 &  0.001 $\pm$ 0.006 \\          
Sb2-b'   &      0.117 & 0.074 $\pm$ 0.006\\
Sb2-b''   &     0.116 & 0.076 $\pm$ 0.006\\
\hline
\hline
\end{tabular}
\end{table}
Taken together, the above results allow us to unambiguously conclude that the kagome planes undergo an ISD distortion in RbV$_{3}$Sb$_{5}$ and support the presence of a $\pi$-shift along $c$ direction, in agreement with previous theoretical predictions \cite{kang2022,  tan2021charge,PhysRevLett.127.236401}.

\begin{figure}[htbp]
\centering
{\includegraphics[width = \linewidth]{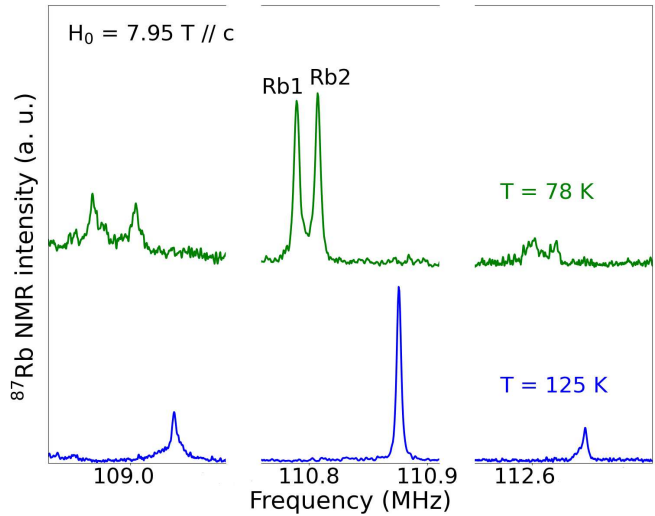}}
\caption{Temperature dependence of the $^{87}$Rb NMR spectra above and below $T_{CDW}$ = 103 K, with $H_{0}$ = 7.95 T // $c$ axis.}
\label{fig:87RbNMR}
\end{figure}

To validate the proposed arrangement of the kagome layers along the $c$ axis, we study the symmetry change of the Rb site.
Fig.~\ref{fig:87RbNMR} shows selected spectra of $^{87}$Rb with $H_{0}$ = 7.95 T // $c$ axis.
As expected for the quadrupole perturbed NMR spectra of the $I$ = 3/2 $^{87}$Rb nuclear spin, three peaks appear above $T_{CDW}$ = 103 K, related to the $\pm1/2$ transition for the central line plus the $+3/2\rightarrow+1/2$ and $-1/2\rightarrow-3/2$ transitions for the low and high-frequency satellites, respectively.
Below $T_{CDW}$ = 103 K, both the satellites and the central line clearly split thus reflecting the presence of two inequivalent sites, labeled Rb1 and Rb2. 

The splitting of all the transition peaks for the $I=3/2$ is due to the consequence of different electric field gradients and local susceptibilities at the two sites, as detailed in the SM~\cite{Note1}. This behavior is in agreement with the behavior of the $^{51}$V NMR in CsV$_{3}$Sb$_{5}$ \cite{song2021orbital,nie2022}.

As already done for $^{121}$Sb NQR measurements, we consider the relative area between the two NMR peaks of a doublet, which is proportional to the ratio between the multiplicity of the corresponding inequivalent atomic sites. 
The stacked ISD layer with a $\pi$-shift \cite{nie2022} yields to a 1:1 multiplicity ratio between Rb1 and Rb2 sites.
On the other hand, the structures with alternating SD and ISD layers \cite{OrtizPhysRevX2021} proposed for CsV$_{3}$Sb${5}$ or stacked of ISD layers without $\pi$-shift yield to a 3:1 ratio \footnote{The comparison among different structures can be seen in Supplementary Materials of Ref.\cite{nie2022}}.
Fig.~\ref{fig:87RbNMR} clearly shows that the ratio between Rb1 and Rb2 peak intensities is nearly equal to 1:1, either for the central or satellite doublets. This result excludes the alternate SD and ISD configuration and is compatible with the stacking made by ISD layers only, $\pi$-shifted from one layer to the other. 

Finally, $^{51}$V NMR results are in agreement with the reports already published for A = Cs \cite{song2021orbital,Mu_2022,nie2022} despite the difference in the spacing layer. Moreover, since the V1:V2 occupation ratio is predicted to be 1:1 for both the SD and ISD structures, this probe is of little interest for our purpose of identifying the CDW configuration of the kagome plane.
These results are reported in the SM~\cite{Note1}.
In conclusion, we measured the zero-field $^{121}$Sb spectrum and the applied field NMR spectra of 
$^{51}$V and $^{87}$Rb.
The analysis of the NMR/NQR spectra of V, Sb, and Rb nuclei, supported by DFT simulations, allows to unambiguously identify the structure stabilized below the CDW transition occurring at $T_{CDW}$ = 103 K.
The analysis of the multiplicity of the nonequivalent sites for each species allows identifying the candidate low-temperature structure: a 2$\times$2$\times$2 superlattices formed by alternating Inverse Star-of-David layers, $\pi$-shifted from one to the other. 

Work at Brown was supported in part by the the National Science Foundation grant No. DMR-1905532 and funds from Brown and University of Bologna.

S.D.W. and B.R.O. gratefully acknowledge support via the UC Santa Barbara NSF Quantum Foundry funded via the Q-AMASE-i program under award DMR-1906325.


\bibliography{bibliography}

\end{document}